\PassOptionsToPackage{hyphens}{url}
\documentclass{iopjournal}

\usepackage{amsmath}
\usepackage{booktabs}
\usepackage{float}
\usepackage{array}
\usepackage{tikz}
\usetikzlibrary{arrows.meta,fit,backgrounds,positioning}
\usepackage{ragged2e}
\usepackage{etoolbox}

\makeatletter
\patchcmd{\@makecaption}{\raggedright}{\justifying}{}{}
\makeatother

\begin{document}
\justifying

\title{Classifying detectability and time of emergence across different Stratospheric Aerosol Injection strategies}

\author{Sai Charan Petchetti$^1$ and Daniele Visioni$^{2,*}$}

\affil{$^1$Department of Physics and Astronomy, University of California, Los Angeles, Los Angeles, CA, USA}

\affil{$^2$Department of Earth and Atmospheric Sciences, Cornell University, Ithaca, NY, USA}

\affil{$^*$Author to whom any correspondence should be addressed.}

\email{\href{mailto:dv224@cornell.edu}{dv224@cornell.edu}}


\begin{abstract}
\justifying\noindent
Stratospheric aerosol injection (SAI) can be deployed using different injection strategies with the same global-mean cooling target and result in distinct regional climate responses. The ability to identify the deployed strategy from observed climate depends on how large the regional differences are relative to internal variability. In this study, we ask whether five feedback-controlled SAI strategies from the GAUSS ensemble (CESM2(WACCM6), SSP2-4.5), all targeting 1.0$^\circ$C above pre-industrial levels, can be distinguished from annual global and regional means of near-surface temperature, precipitation, and polar sea-ice fraction, and when this distinguishability emerges. We use multinomial logistic-regression classifiers with leave-one-member-out cross-validation and define the emergence year using a family-wise error rate threshold computed from a restricted-permutation null, combined with a calibrated persistence requirement. We find that the strategies become distinguishable in the mid-to-late 2040s, about 8--14 years after deployment, with a peak accuracy of 0.87 on a model using both global and regional features. Additionally, we find that no model trained on a single region--variable pair performs as well as the models using global and regional means, suggesting the discriminative signal is small and distributed across regions and variables. The equatorial and high-latitude injection strategies are the most distinguishable. The classifier is able to recognize the multi-latitude strategy more easily at lower temperature targets. Our results suggest that annual climate features can identify with confidence which coordinated injection strategy has been deployed roughly a decade or more after deployment.
\end{abstract}

\section{Introduction}
Global efforts to reduce greenhouse gas emissions are unlikely, on their own, to avoid some of the impacts of climate change \cite{unep2026overshoot}, especially if carbon dioxide removal (CDR) methods will not scale up as hoped. This has motivated research into climate intervention methods such as stratospheric aerosol injection (SAI) as a possible supplement to emissions reduction and other mitigation strategies \cite{macmartin2022} to avoid temperature overshoots in the coming decades. An SAI deployment scenario would be subject to a number of choices such as the target cooling, the date of the start of a deployment, and the injection strategy itself.

It has been well established in the literature that different injection strategies can lead to the same global-mean cooling but distinct regional outcomes. Equatorial injection tends to result in residual overcooling in the tropics and undercooling at the poles \cite{kravitz2019}. Injecting elsewhere can either maintain hemispheric symmetry, using a single latitude pair \cite{zhang2024, bednarz2023}, or follow a more complex multi-latitude, feedback-controlled design, such as that of Kravitz et al. \cite{kravitz2017}, which simultaneously balances global-mean temperature and the interhemispheric and Equator-to-pole temperature gradients. Other strategies with different targets  and different trade-offs are also possible, such as those with precipitation-based targets \cite{lee2020}. Zhang et al. \cite{zhang2024} and Bednarz et al. \cite{bednarz2023} compare these symmetric and multi-objective strategies under a common scenario and find that the choice of strategy impacts aerosol distribution, injection efficiency, and a wide range of surface responses, even when all strategies target the same global-mean temperature. As deployment need not be coordinated by a single actor \cite{maattanen2026} and different strategies have different regional trade-offs \cite{zhang2024}, being able to tell which strategy is deployed from the observed climate response is relevant for monitoring, attribution, and governance. A useful framework to address this question is the time of emergence, an established methodology in climate research to estimate when a forced signal becomes distinguishable from internal variability \cite{hawkins2012}, and more recently studied using machine learning models \cite{barnes2020}.

A novel line of work has used machine-learning classifiers to quantify the detectability of SAI’s climate signal, but has so far addressed easier questions with higher class separability than strategy attribution. Mamalakis et al. \cite{mamalakis2023}, for instance, trained a neural network to quantify how distinguishable a post-deployment SAI climate might be from a pre-deployment climate. Similarly, Labe et al. \cite{labe2023} use the same model, background scenario and deployment year as this study, and trained a logistic-regression classifier to detect whether a single SAI strategy has been deployed, year by year, finding the signal detectable within about a decade. Meanwhile, Dong et al. \cite{dong2026} classified which of seven \emph{unilateral}, uncoordinated injection latitudes caused a given climate response, and find them distinguishable within about two years. None of these studies ask which of several \emph{coordinated} strategies -- all independently targeting the same global-mean temperature -- is responsible for an observed climate, nor do they provide a significance framework that controls for testing many candidate predictors and many candidate years simultaneously.

In this study, we address both gaps. We train a multinomial logistic-regression classifier to distinguish between five feedback-controlled SAI strategies from the Geoengineering Assessment across Uncertainty, Scenarios, and Strategies (GAUSS) dataset, all targeting the same $1.0^\circ$C-above-pre-industrial temperature under a common background scenario, using annual regional and global means of near-surface temperature, precipitation and polar sea-ice fraction. We assess classifier performance by leave-one-member-out cross-validation, tested against a restricted-permutation null that preserves each simulation's temporal structure. This null yields a family-wise significance threshold and a calibrated run-length criterion that together define an emergence year. This framework allows us to quantify whether coordinated strategies are distinguishable in addition to when and which regions and variables carry that signal. We also test whether the trained classifier generalizes beyond the conditions it was trained on by applying it to multi-latitude simulations at two other temperature targets.

\section{Dataset}
\label{sec:dataset}

This work uses the GAUSS dataset \cite{macmartin2022}. GAUSS contains a set of SAI simulation experiments run in the middle-atmosphere configuration of CESM2(WACCM6) \cite{danabasoglu2020, gettelman2019} that aim to maintain temperatures at various global warming levels relative to pre-industrial levels, using the Shared Socioeconomic Pathway (SSP)2-4.5 as the background emissions scenario. We use simulations that target a global-mean temperature 1.0$^\circ$C above pre-industrial levels previously described in \cite{zhang2024}. In these simulations, the temperature target ramps down gradually from about 1.5$^\circ$C in 2035 at the start of deployment to 1.0$^\circ$C over the first ten years \cite{zhang2024}.

We compare five different injection strategies: single-point injection at the equator (0N), symmetric injections at 15N+15S, 30N+30S, and 60N+60S, and a multi-latitude feedback-controlled strategy. Each strategy simulation contains three ensemble members. For the purposes of this study, we use the monthly output from 2035--2069 for near-surface air temperature, precipitation, and polar sea-ice fraction, averaged to yearly values (for sea-ice fraction, September and March values only). To test generalizability of the detection signal, we include additional GAUSS multi-latitude simulations at 0.5$^\circ$C and 1.5$^\circ$C above preindustrial temperatures (three ensemble members for each temperature target).

\section{Method}

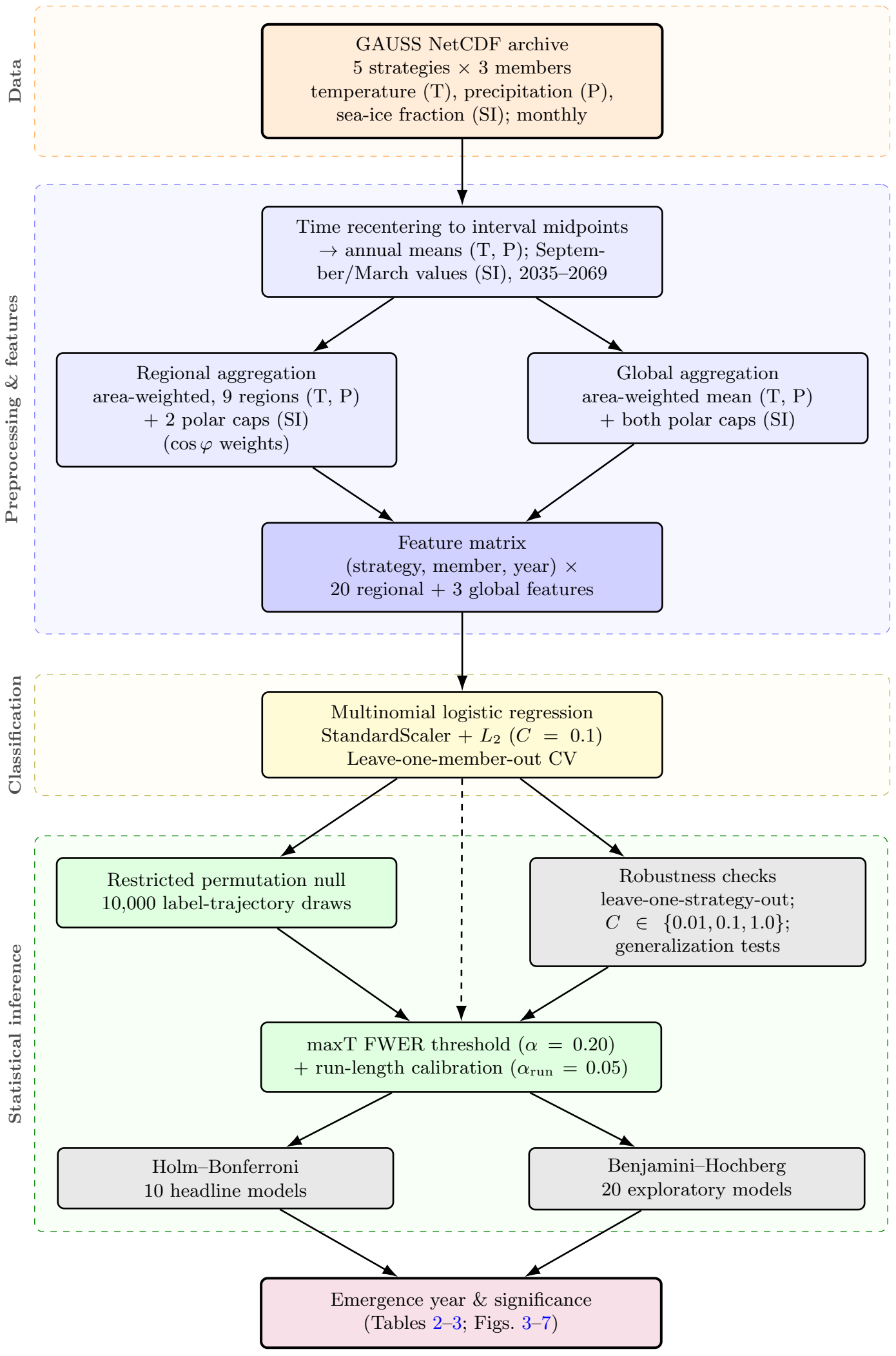
\begin{figure}[tbp]
\centering
\scalebox{0.62}{
\begin{tikzpicture}[
  font=\small,
  node distance=0.9cm,
  block/.style={rectangle, rounded corners=3pt, draw=black, thick, fill=blue!8,
                text width=5.2cm, minimum height=1.15cm, align=center, inner sep=6pt},
  datablock/.style={rectangle, rounded corners=3pt, draw=black, very thick, fill=orange!15,
                text width=5.2cm, minimum height=1.15cm, align=center, inner sep=6pt},
  procblock/.style={rectangle, rounded corners=3pt, draw=black, thick, fill=yellow!20,
                text width=5.2cm, minimum height=1.15cm, align=center, inner sep=6pt},
  statblock/.style={rectangle, rounded corners=3pt, draw=black, thick, fill=green!12,
                text width=5.2cm, minimum height=1.15cm, align=center, inner sep=6pt},
  corrblock/.style={rectangle, rounded corners=3pt, draw=black, thick, fill=gray!18,
                text width=5.2cm, minimum height=1.0cm, align=center, inner sep=5pt},
  outblock/.style={rectangle, rounded corners=3pt, draw=black, very thick, fill=purple!12,
                text width=6.2cm, minimum height=1.15cm, align=center, inner sep=6pt},
  row/.style={inner sep=0pt, outer sep=0pt},  
  arrow/.style={-{Latex[length=2.6mm, width=1.8mm]}, thick},
  layerlabel/.style={font=\footnotesize\bfseries, text=gray!60!black, anchor=south, rotate=90}
]

\node[datablock, text width=6.2cm] (netcdf)
  {GAUSS NetCDF archive\\5 strategies $\times$ 3 members\\temperature (T), precipitation (P),\\sea-ice fraction (SI); monthly};

\node[block, text width=6.2cm, below=1.1cm of netcdf] (preproc)
  {Time recentering to interval midpoints $\rightarrow$ annual means (T, P); September/March values (SI), 2035--2069};

\node[block, anchor=north] (regional) at ([xshift=-3.9cm, yshift=-0.9cm]preproc.south)
  {Regional aggregation\\area-weighted, 9 regions (T, P)\\+ 2 polar caps (SI)\\($\cos\varphi$ weights)};
\node[block, anchor=north] (global) at ([xshift=3.9cm, yshift=-0.9cm]preproc.south)
  {Global aggregation\\area-weighted mean (T, P)\\+ both polar caps (SI)};
\node[row, fit=(regional)(global)] (aggrow) {};

\node[block, fill=blue!18, text width=6.2cm, below=of aggrow] (featmat)
  {Feature matrix\\(strategy, member, year) $\times$\\20 regional + 3 global features};

\node[procblock, text width=6.2cm, below=1.3cm of featmat] (classifier)
  {Multinomial logistic regression\\StandardScaler + $L_2$ ($C=0.1$)\\Leave-one-member-out CV};

\node[statblock, anchor=north] (nullperm) at ([xshift=-3.9cm, yshift=-1.3cm]classifier.south)
  {Restricted permutation null\\10{,}000 label-trajectory draws};
\node[corrblock, anchor=north] (robust) at ([xshift=3.9cm, yshift=-1.3cm]classifier.south)
  {Robustness checks\\leave-one-strategy-out;\\$C \in \{0.01,0.1,1.0\}$;\\generalization tests};
\node[row, fit=(nullperm)(robust)] (testrow) {};

\node[statblock, text width=6.2cm, below=of testrow] (fwer)
  {maxT FWER threshold ($\alpha=0.20$)\\+ run-length calibration ($\alpha_{\mathrm{run}}=0.05$)};

\node[corrblock, anchor=north] (holm) at ([xshift=-3.9cm, yshift=-0.9cm]fwer.south)
  {Holm--Bonferroni\\10 headline models};
\node[corrblock, anchor=north] (bh) at ([xshift=3.9cm, yshift=-0.9cm]fwer.south)
  {Benjamini--Hochberg\\20 exploratory models};
\node[row, fit=(holm)(bh)] (corrrow) {};

\node[outblock, below=1.1cm of corrrow] (emergence)
  {Emergence year \& significance\\(Tables~\ref{tab:headline}--\ref{tab:csensitivity}; Figs.~\ref{fig:exploratory21}--\ref{fig:generalization_class})};

\node[row, fit=(regional)(global)(nullperm)(robust)(holm)(bh)] (allwidth) {};

\begin{scope}[on background layer]
\node[fit=(netcdf)(allwidth.west |- netcdf)(allwidth.east |- netcdf), inner xsep=10pt, inner ysep=8pt, draw=orange!60, dashed, rounded corners, fill=orange!3] (g0) {};
\node[fit=(preproc)(aggrow)(featmat), inner sep=10pt, draw=blue!50, dashed, rounded corners, fill=blue!3] (g1) {};
\node[fit=(classifier)(allwidth.west |- classifier)(allwidth.east |- classifier), inner xsep=10pt, inner ysep=8pt, draw=yellow!70!black, dashed, rounded corners, fill=yellow!3] (g2) {};
\node[fit=(testrow)(fwer)(corrrow), inner sep=10pt, draw=green!50!black, dashed, rounded corners, fill=green!3] (g3) {};
\end{scope}

\node[layerlabel] at ([xshift=-0.1cm]g0.west) {Data};
\node[layerlabel] at ([xshift=-0.1cm]g1.west) {Preprocessing \& features};
\node[layerlabel] at ([xshift=-0.1cm]g2.west) {Classification};
\node[layerlabel] at ([xshift=-0.1cm]g3.west) {Statistical inference};

\draw[arrow] (netcdf) -- (preproc);
\draw[arrow] (preproc) -- (regional);
\draw[arrow] (preproc) -- (global);
\draw[arrow] (regional) -- (featmat);
\draw[arrow] (global) -- (featmat);
\draw[arrow] (featmat) -- (classifier);
\draw[arrow] (classifier) -- (nullperm);
\draw[arrow] (classifier) -- (robust);
\draw[arrow, dashed] (classifier) -- (fwer);
\draw[arrow] (nullperm) -- (fwer);
\draw[arrow] (robust) -- (fwer);
\draw[arrow] (fwer) -- (holm);
\draw[arrow] (fwer) -- (bh);
\draw[arrow] (holm) -- (emergence);
\draw[arrow] (bh) -- (emergence);

\end{tikzpicture}

}
\caption{Systems architecture of the classification and emergence-detection pipeline. The dashed arrow represents the empirical accuracy curve fed into the significance test along with the null distribution.}
\label{fig:architecture}
\end{figure}

For each ensemble member, we computed annual regional and global means of near-surface air temperature and precipitation, and polar means of September (for Arctic) and March (for Antarctic) sea-ice fraction (Figure~\ref{fig:architecture}). We re-centered monthly output. In the case of the GAUSS multi-latitude strategy ensemble members at all three temperature targets, for which time bounds were not recorded, a 15-day backward shift was implemented. This re-centering ensures that each calendar-year monthly series covers January-December and that the full 2035-2069 period is retained.

\begin{figure}[H]
\centering
\includegraphics[width=\textwidth]{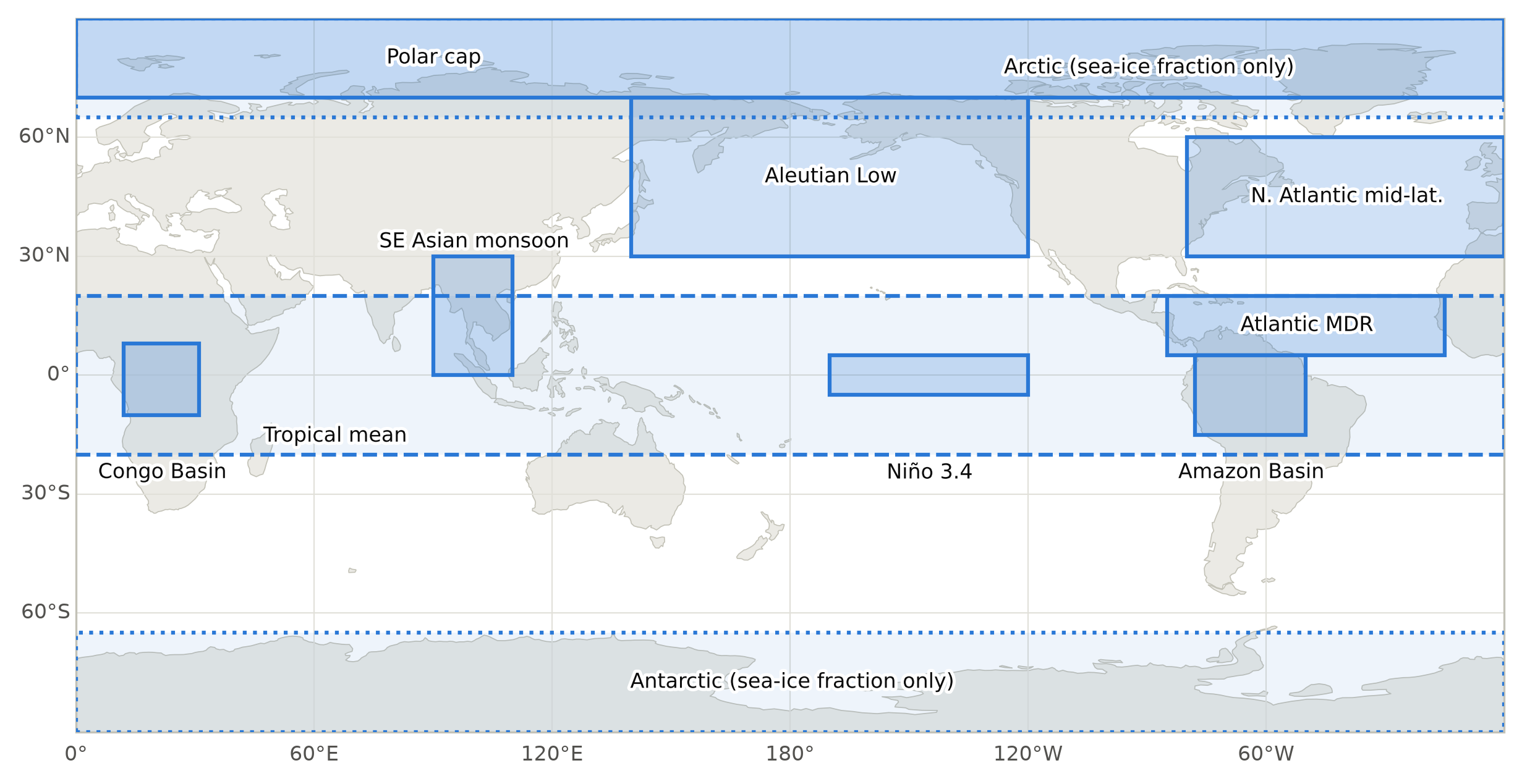}
\caption{Regions used for the regional features. Temperature and precipitation are averaged over the nine regions (solid outlines; the tropical-mean band is dashed). Sea-ice fraction is averaged over the Arctic (65$^\circ$N--90$^\circ$N, September) and Antarctic (65$^\circ$S--90$^\circ$S, March) caps (dotted outlines).}
\label{fig:regions}
\end{figure}

We calculate and use both global and regional averages. The regions are defined as follows (Figure~\ref{fig:regions}):

\begin{itemize}
    \item Tropical mean: 20$^\circ$S--20$^\circ$N, all longitudes \cite{zhang2024}
    \item Amazon Basin: 15$^\circ$S--5$^\circ$N, 78$^\circ$W--50$^\circ$W \cite{zhang2024}
    \item Congo Basin: 10$^\circ$S--8$^\circ$N, 12$^\circ$E--31$^\circ$E \cite{zhang2024}
    \item Ni\~{n}o 3.4: 5$^\circ$S--5$^\circ$N, 170$^\circ$W--120$^\circ$W \cite{zhang2024}
    \item Atlantic Main Development Region (MDR): 5$^\circ$N--20$^\circ$N, 85$^\circ$W--15$^\circ$W \cite{zhang2024}
    \item North Atlantic mid-latitudes: 30$^\circ$N--60$^\circ$N, 80$^\circ$W--0$^\circ$ \cite{bednarz2023}
    \item Polar cap: 70$^\circ$N--90$^\circ$N, all longitudes \cite{bednarz2023}
    \item Aleutian Low region: 30$^\circ$N--70$^\circ$N, 140$^\circ$E--120$^\circ$W \cite{bednarz2023}
    \item Southeast Asian monsoon region: 0$^\circ$--30$^\circ$N, 90$^\circ$E--110$^\circ$E \cite{narenpitak2024}
    \item Arctic: 65$^\circ$N--90$^\circ$N, all longitudes (September sea-ice fraction only)
    \item Antarctic: 65$^\circ$S--90$^\circ$S, all longitudes (March sea-ice fraction only)
\end{itemize}

The region definitions were motivated by literature suggesting these were some of the regions that might be the most impacted by SAI-driven changes. Zhang et al. \cite{zhang2024} found that the closer the injection of SO$_2$ is to the equator, the stronger the decrease in tropical mean precipitation per degree of tropical cooling. They found that the equatorial strategy results in the largest reduction, followed by 15N+15S, 30N+30S, and 60N+60S in that order. We use the same tropical mean region defined by Zhang et al. \cite{zhang2024}, aiming to capture similar discriminative power between strategies. Similarly, we use the Amazon Basin, Congo Basin, Ni\~{n}o 3.4, and Atlantic MDR regions from Zhang et al. \cite{zhang2024}, who identify each region as particularly sensitive to the choice of injection strategy due to shifts associated with tropical circulation and El Ni\~{n}o-Southern Oscillation (ENSO)-related variability. For the Amazon and Congo basins, Simpson et al. \cite{simpson2019} analysed the hydroclimate sensitivity to SAI. In the Congo basin, Abiodun et al. \cite{abiodun2021} and Obahoundje et al. \cite{obahoundje2023} demonstrate that SAI deployment alters drought risk and compound precipitation-temperature extremes. The Ni\~{n}o 3.4 region is usually used to monitor ENSO \cite{trenberth1997}. While Gabriel and Robock \cite{gabriel2015} find that SAI does not significantly alter ENSO frequency or amplitude, the tropical Pacific mean-state response they document remains dependent on the applied forcing, supporting the use of the Ni\~{n}o 3.4 region in our study. The Atlantic MDR region, originally defined as the primary origin region for major Atlantic hurricanes \cite{goldenberg2001}, has subsequently been shown to be sensitive to SAI deployment \cite{moore2015}. Furthermore, we make use of the North Atlantic mid-latitude, polar cap and Aleutian Low regions from Bednarz et al. \cite{bednarz2023}, who use these to construct winter sea-level-pressure indices that show a response connecting injection latitude to Northern Hemisphere surface climate. Finally, we use the Southeast Asian monsoon region from Narenpitak et al. \cite{narenpitak2024}, who demonstrate that this monsoon-dependent region is particularly sensitive to precipitation changes under both greenhouse gas emission scenarios and different solar radiation modification (SRM) approaches.

Area-weighted annual means were calculated over the nine regions using weights proportional to the cosine of the latitude of each grid-cell center. Sea-ice fraction was calculated only in the Arctic and Antarctic regions, for the months of September and March respectively, leaving 20 region--variable features. The Arctic and Antarctic sea-ice features are used separately as regional features and together as the global sea-ice feature.

Multinomial logistic regression classifiers ($L_{2}$ penalty, inverse strength $C = 0.1$), with input features standardized to zero mean and unit variance on each training fold, were trained to identify the injection strategy from yearly feature averages. The models were evaluated using leave-one-member-out cross-validation: each of three folds was trained on two ensemble members and tested on the held-out member, with classification accuracy scored separately for each calendar year. For example, when ensemble member~001 is held out, the model is trained on members~002 and~003 (all five strategies) and tested on the five strategies' yearly data from member~001. Since there are only five test points per fold-year, a single fold-year accuracy takes one of six values. To help stabilize this small-sample noise, the three fold accuracies are averaged per calendar year, and a 5-year centred running mean of the resulting curve is also computed. The emergence year aims to capture the first year at which the strategies can be reliably distinguished. A naive emergence year is calculated from the 5-year smoothed mean accuracy, defined as the first year the smoothed accuracy exceeds 0.5. A run-based emergence year is also calculated from the unsmoothed mean-accuracy curve, using a significance threshold derived from a 10{,}000-draw null permutation test.

Each of the 10{,}000 null draws applies a bijection of the five labels to entire trajectories -- all 35 years move together to a new strategy label -- drawn independently for each ensemble member. The maximum-statistic (maxT) permutation test is applied at the $\alpha = 0.20$ level: the threshold is set to the 80th percentile of the distribution of the null curves' maximum accuracy across years. The emergence year is then identified by requiring a run of consecutive above-threshold years long enough that runs of that length appear in at most 5\% of null curves ($\alpha_{\mathrm{run}} = 0.05$). In the main analysis, this calibrated run length was two years for all headline models. Emergence is therefore declared at the first year from which the unsmoothed fold-averaged accuracy exceeds the 80th-percentile threshold for two consecutive years. A Monte Carlo $p$-value with the $+1$ correction is then computed and reported. The $+1$ correction adds one to both the numerator and denominator to avoid zero $p$-values.

The Holm-Bonferroni method is applied across ten pre-specified headline models to control the family-wise error rate (FWER). The ten headline models are: the regional temperature and precipitation models, the all-variable regional model, global-only, global-plus-regional, the three single global-variable models, and the Arctic and Antarctic sea-ice models. The less conservative Benjamini-Hochberg method is applied across 20 single region--variable exploratory models to control the false discovery rate (FDR) ($2\ \text{variables} \times 9\ \text{regions} + 2$ polar sea-ice fraction features). Robustness was assessed by leaving out each of the strategies (10{,}000 null permutations) and by varying $C$ over $\{0.01,\ 0.1,\ 1.0\}$ (1{,}000 null permutations).

\begin{table}[!tb]
\centering
\caption{Design of the main analysis and the cross-target generalization test.}
\label{tab:design}
\small
\begin{tabular}{>{\raggedright\arraybackslash}p{2.1cm}>{\raggedright\arraybackslash}p{4.4cm}>{\raggedright\arraybackslash}p{4.4cm}}
\toprule
 & Main analysis & Cross-target \\
\midrule
Question & Which of the five strategies is deployed?
& Is the multi-latitude strategy still recognized at a
different temperature target? \\
\addlinespace
Training data & \multicolumn{2}{>{\raggedright\arraybackslash}p{8.8cm}}{GAUSS,
1.0$^\circ$C target, all five strategies, two of three
members (leave-one-member-out)} \\
\addlinespace
Test data & Held-out member, all five strategies,
1.0$^\circ$C & Held-out member's multi-latitude
simulation at 1.0$^\circ$C (baseline), 0.5$^\circ$C,
and 1.5$^\circ$C \\
\addlinespace
Yearly score & Accuracy (fraction of strategies
correctly classified) & Probability assigned to the
multi-latitude strategy \\
\addlinespace
Averaged over & Three folds & Three folds \\
\addlinespace
Null: training labels & \multicolumn{2}{>{\raggedright\arraybackslash}p{8.8cm}}{Strategy
labels of simulations shuffled within each ensemble
member (10{,}000 draws)} \\
\addlinespace
Null: test labels & Shuffled with the held-out
member's own labels & Each simulation takes its
member's shuffled multi-latitude label \\
\addlinespace
Significance & \multicolumn{2}{>{\raggedright\arraybackslash}p{8.8cm}}{Peak
$p$-value: fraction of null curves whose highest yearly
score is at least as high as the real curve's highest
yearly score ($+1$ correction)} \\
\addlinespace
Run length for emergence & 2 years & 2--5 years
(calibrated per test) \\
\addlinespace
Correction across models & Holm--Bonferroni (10
headline), Benjamini--Hochberg (20 exploratory) &
None \\
\bottomrule
\end{tabular}
\end{table}

In the generalization tests, the classifiers are trained on the 1.0$^\circ$C GAUSS simulations and tested on multi-latitude simulations they were not trained on (Table~\ref{tab:design}). Each classifier is tested on the multi-latitude simulation of its held-out ensemble member at the 1.0$^\circ$C (training target, as a baseline), 0.5$^\circ$C, and 1.5$^\circ$C targets. For example, the classifier trained on members~002 and~003 is tested on member~001 at each temperature target.

The generalization tests are evaluated only on the multi-latitude strategy. Since there are only three GAUSS ensemble members, the fraction of multi-latitude GAUSS simulations correctly classified in each year can only take the values 0, 1/3, 2/3, or 1, which is too coarse for a significance test. To account for this, the five-strategy classifier trained on the 1.0$^\circ$C GAUSS dataset is tested on the generalization test datasets and each year is scored by the probability the classifier assigns to the multi-latitude strategy, averaged over the three folds. In addition to this, a significance test is conducted on 10{,}000 null permutation draws. In each draw, the classifiers are retrained on the 1.0$^\circ$C GAUSS simulations with strategy labels shuffled within each ensemble member, as in the main analysis, and tested on the generalization test datasets.  The yearly scores of the classifiers trained on the unshuffled labels are then compared against the 10,000 null score curves. The peak $p$-value is the fraction of draws whose highest yearly score is at least as high as the highest yearly score of the unshuffled classifiers, with the $+1$ correction. The generalization test results are not Holm-Bonferroni corrected. 

Emergence years are computed as in the main analysis, but the calibrated run length is often longer and differs between tests (between two and five years). Since the test datasets contain only multi-latitude simulations, each null draw scores a smoothly varying probability for a single strategy, rather than an accuracy over all five, so null curves that cross the threshold tend to stay above it for many consecutive years. These emergence years are therefore generally more conservative than, and not directly comparable to, those of the main analysis.

\section{Results}

\begin{table}[H]
\centering
\caption{Headline classification results for the ten pre-specified feature sets (5 strategies, chance $=0.20$). EY denotes emergence year. Rows are ordered by peak permutation $p$-value; the final column reports significance under Holm--Bonferroni control of the FWER at $\alpha=0.05$. Holm--Bonferroni correction applies to peak $p$-values only; $^\dagger$ marks emergence years of models that fail Holm--Bonferroni.
\newline
}
\label{tab:headline}
\begin{tabular}{lccccc}
\toprule
Feature set & Naive EY & Run-based EY & Peak acc. & Peak $p$ & Holm sig. \\
\midrule
Global + regional (all vars) & 2043 & 2043 & 0.87 & 0.0006 & Yes \\
All regional (all vars)      & 2044 & 2045 & 0.80 & 0.0009 & Yes \\
Global only (all vars)       & 2047 & 2049 & 0.73 & 0.0028 & Yes \\
Sea-ice fraction, Antarctic  & --   & 2063 & 0.60 & 0.0068 & Yes \\
Precipitation, regional      & 2051 & 2055$^\dagger$ & 0.67 & 0.0127 & No  \\
Temperature, regional        & 2053 & 2050$^\dagger$ & 0.67 & 0.0132 & No  \\
Precipitation, global        & --   & 2049$^\dagger$ & 0.53 & 0.0366 & No  \\
Sea-ice fraction, global & --   & 2059$^\dagger$ & 0.53 & 0.0926 & No  \\
Sea-ice fraction, Arctic     & --   & --   & 0.40 & 0.4084 & No  \\
Temperature, global          & --   & --   & 0.40 & 0.4675 & No  \\
\bottomrule
\end{tabular}
\end{table}

Table~\ref{tab:headline} presents results for the ten pre-specified headline models. Peak $p$-values for the three all-variable headline models (global-plus-regional, all-regional, and global-only) and the Antarctic sea-ice model remain significant after Holm--Bonferroni correction at $\alpha = 0.05$. The regional precipitation and temperature, and global precipitation and sea-ice models emerge between 2049 and 2059, but do not survive the FWER correction. The Arctic sea-ice and global temperature models do not emerge and do not survive FWER correction.

The above results for global-mean temperature reflect the design of the GAUSS experiments, which include a feedback mechanism to vary injection amounts annually to hold the global-mean surface temperature fixed for all injection strategies. In addition to the global-mean surface temperature, the multi-latitude injection strategy was designed to hold interhemispheric and equator-to-pole gradients constant. Of the global-mean properties, surface temperature is, as expected, the least discriminative feature ($p = 0.47$; Table~\ref{tab:headline}). The other global variables, sea-ice fraction and precipitation, carry more discriminative signal than global temperature (peak accuracies of 0.53 for both, compared to 0.40) but do not survive FWER correction.

\begin{figure}[H]
\centering
\includegraphics[width=\textwidth]{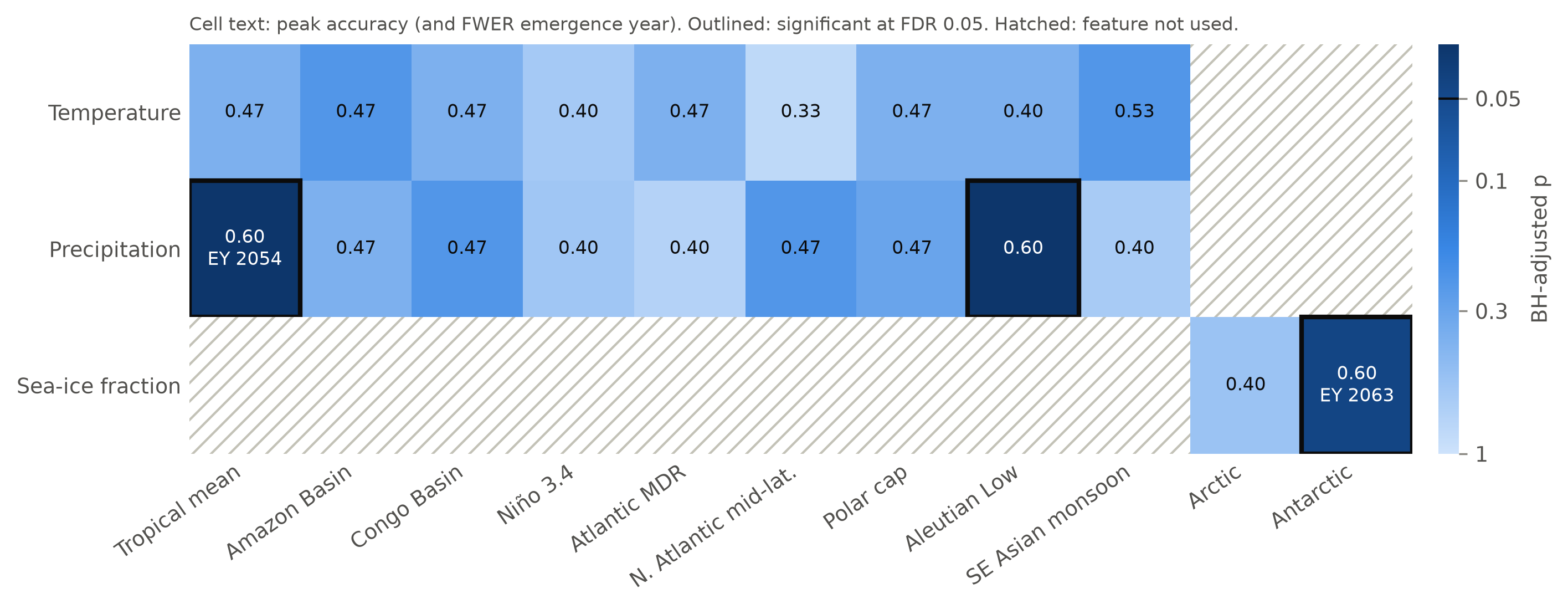}
\caption{Classification results of the 20 exploratory single region $\times$ variable models (5 strategies, chance $=0.20$). Cell color indicates the Benjamini--Hochberg adjusted peak $p$-value; cell text gives the peak accuracy and, where one exists, the run-based emergence year. Outlined cells are significant at FDR $=0.05$. Hatched cells are region--variable combinations not used. The naive running-mean accuracy never exceeded 0.5 for any single feature, so no naive emergence year is reported. The ``FWER emergence year'' in the figure header refers to run-based emergence years.}
\label{fig:exploratory21}
\end{figure}

\begin{figure}[H]
\centering
\includegraphics[width=\textwidth]{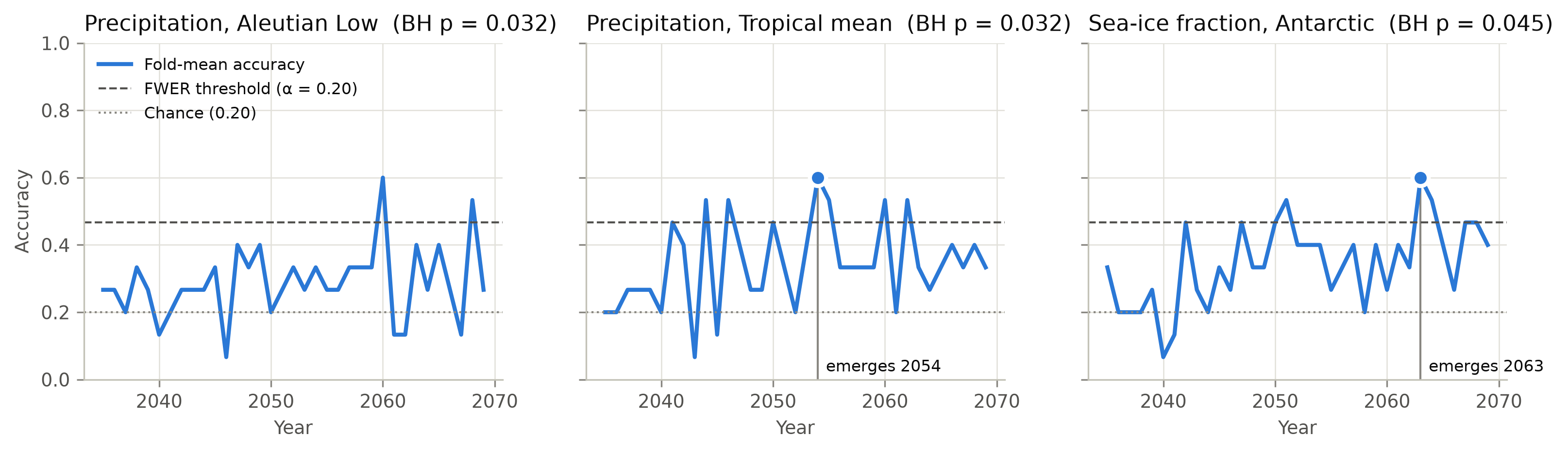}
\caption{Yearly accuracy averaged over the three folds of the three single region $\times$ variable models that survive FDR control. The dashed line is the FWER threshold ($\alpha=0.20$) and the dotted line is chance ($0.20$). The marker indicates the run-based emergence year, the first year of two consecutive years above the threshold. Aleutian-Low precipitation crosses the threshold but never for two consecutive years, so it does not report an emergence year.}
\label{fig:significant21}
\end{figure}

Of the 20 exploratory single region--variable models in Figure~\ref{fig:exploratory21}, only three models survive FDR control (Figure~\ref{fig:significant21}), two of which are precipitation models and one Antarctic sea-ice fraction, and none of the regional temperature models survive FDR control. This is notable given the higher year-to-year variability of precipitation, and may reflect strategy-driven changes in circulation and the intertropical convergence zone (ITCZ) \cite{zhang2024, bednarz2023, lee2020}, although three significant models provide limited evidence. More notably, no single feature performs as well as the combined headline models: the highest peak accuracy achieved by a single feature is 0.60, compared with 0.87 for the global-plus-regional model in Table~\ref{tab:headline}, and no single feature's smoothed accuracy exceeds 0.5. This suggests that the inter-strategy discriminative signal is small and distributed across features, which is the expected outcome of a strategy design that minimizes gross climate differences between strategies.

\begin{figure}[H]
\centering
\includegraphics[width=\textwidth]{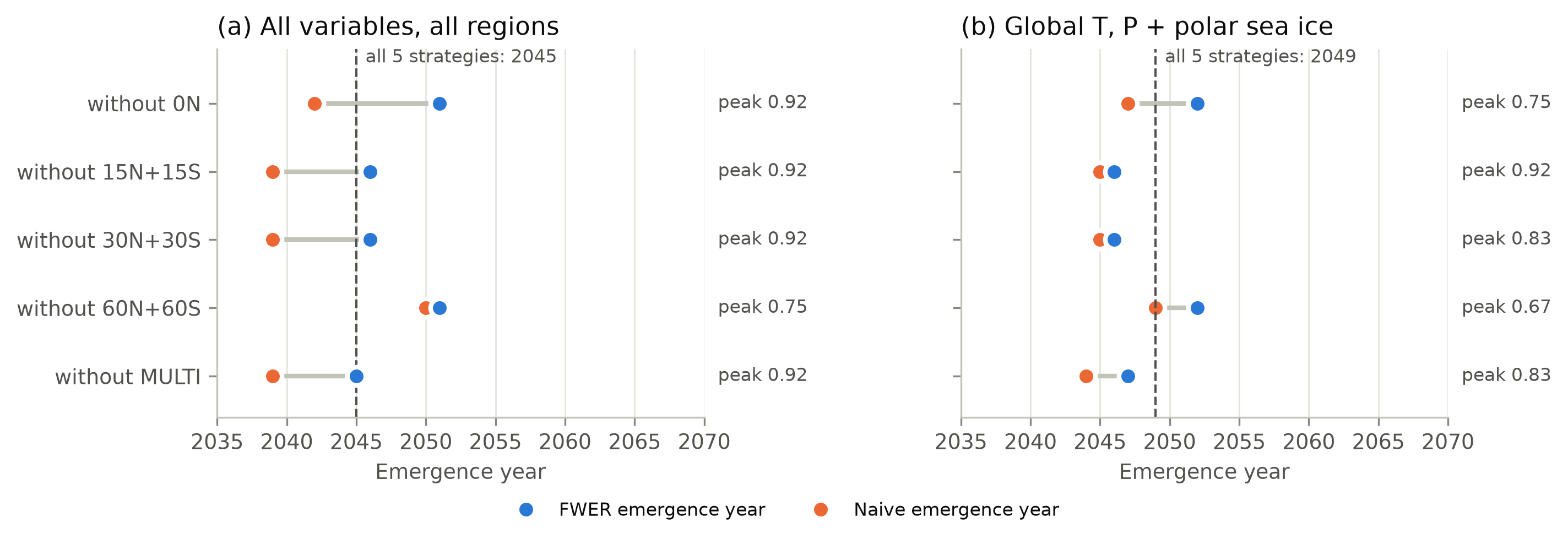}
\caption{Leave-one-strategy-out robustness check. Each strategy is removed in turn and the classifier is retrained on the remaining four (chance $=0.25$). Panel (a) uses all variables and all regions; panel (b) uses the three global variables. Markers show the naive and run-based emergence years for each left-out strategy, and the text on the right gives the peak accuracy. The dashed line marks the run-based emergence year of the corresponding five-strategy model. Nine of the ten models are significant (peak $p \le 0.033$) against their permutation null. The global-only model without 60N+60S is the only model that is not ($p = 0.092$). The ``FWER emergence year'' in the legend refers to run-based emergence years.}
\label{fig:loso}
\end{figure}

From the leave-one-strategy-out analysis shown in Figure~\ref{fig:loso}, we find that the discriminative signal emerges even after any one strategy is left out in both the all-variable all-region models (without global temperature and precipitation) and the global-only models. Leaving out the 0N or 60N+60S strategies delays emergence to 2051--2052, and leaving out 60N+60S also lowers peak accuracy, while leaving out the 15N+15S, 30N+30S or multi-latitude strategies changes the all-regional emergence year by at most one year and brings the global-only emergence forward. 

These results suggest that the equatorial (0N) and high-latitude (60N+60S) strategies are the most distinguishable ones, while the lower-latitude hemispherically symmetric strategies (15N+15S, 30N+30S) and the multi-latitude strategy are the least distinguishable, consistent with the findings of Zhang et al. \cite{zhang2024}, and expected since these two strategies are at the opposite ends of the range of injection latitudes tested, producing climate impacts that are most different from other strategies. Equatorial (0N) strategies largely contain aerosols to the tropics, causing the largest reduction in precipitation per unit of global cooling and undercompensating the Equator-to-pole temperature gradient. The 60N+60S injection strategy, on the other hand, causes the smallest precipitation reduction per unit of global cooling, overcools the Arctic, and undercools the tropics \cite{zhang2024}.

\begin{table}[H]
\centering
\caption{Sensitivity of the logistic-regression models to inverse regularisation strength $C$. Four feature sets are each rerun at $C\in\{0.01,0.1,1.0\}$ (chance $=0.20$; $n_{\mathrm{perm}}=1000$; $p$-values uncorrected).
\newline
}
\label{tab:csensitivity}
\begin{tabular}{lcccc}
\toprule
Feature set & $C$ & Run-based EY & Peak acc. & Peak $p$ \\
\midrule
Global only          & 0.01 & 2052 & 0.67 & 0.0040 \\
Global only          & 0.10 & 2049 & 0.73 & 0.0020 \\
Global only          & 1.00 & 2049 & 0.73 & 0.0040 \\
\addlinespace
Global + regional    & 0.01 & 2049 & 0.80 & 0.0010 \\
Global + regional    & 0.10 & 2043 & 0.87 & 0.0010 \\
Global + regional    & 1.00 & 2039 & 0.93 & 0.0010 \\
\addlinespace
All regional         & 0.01 & 2051 & 0.73 & 0.0060 \\
All regional         & 0.10 & 2045 & 0.80 & 0.0010 \\
All regional         & 1.00 & 2039 & 0.87 & 0.0010 \\
\addlinespace
Precipitation, regional      & 0.01 & 2055 & 0.73 & 0.0030 \\
Precipitation, regional      & 0.10 & 2055 & 0.67 & 0.0200 \\
Precipitation, regional      & 1.00 & 2049 & 0.73 & 0.0060 \\
\bottomrule
\end{tabular}
\end{table}

The regularisation strength sensitivity test (Table~\ref{tab:csensitivity}) confirms that the separability found above is not driven by the chosen regularisation strength, although the emergence year is sensitive to it. For $C \in \{0.01, 0.1, 1.0\}$ the emergence year shifts by up to about six years relative to $C = 0.1$ and every configuration remains significant at $p < 0.05$. Weaker regularisation moves emergence earlier.

\begin{figure}[tbp]
\centering
\includegraphics[width=\textwidth]{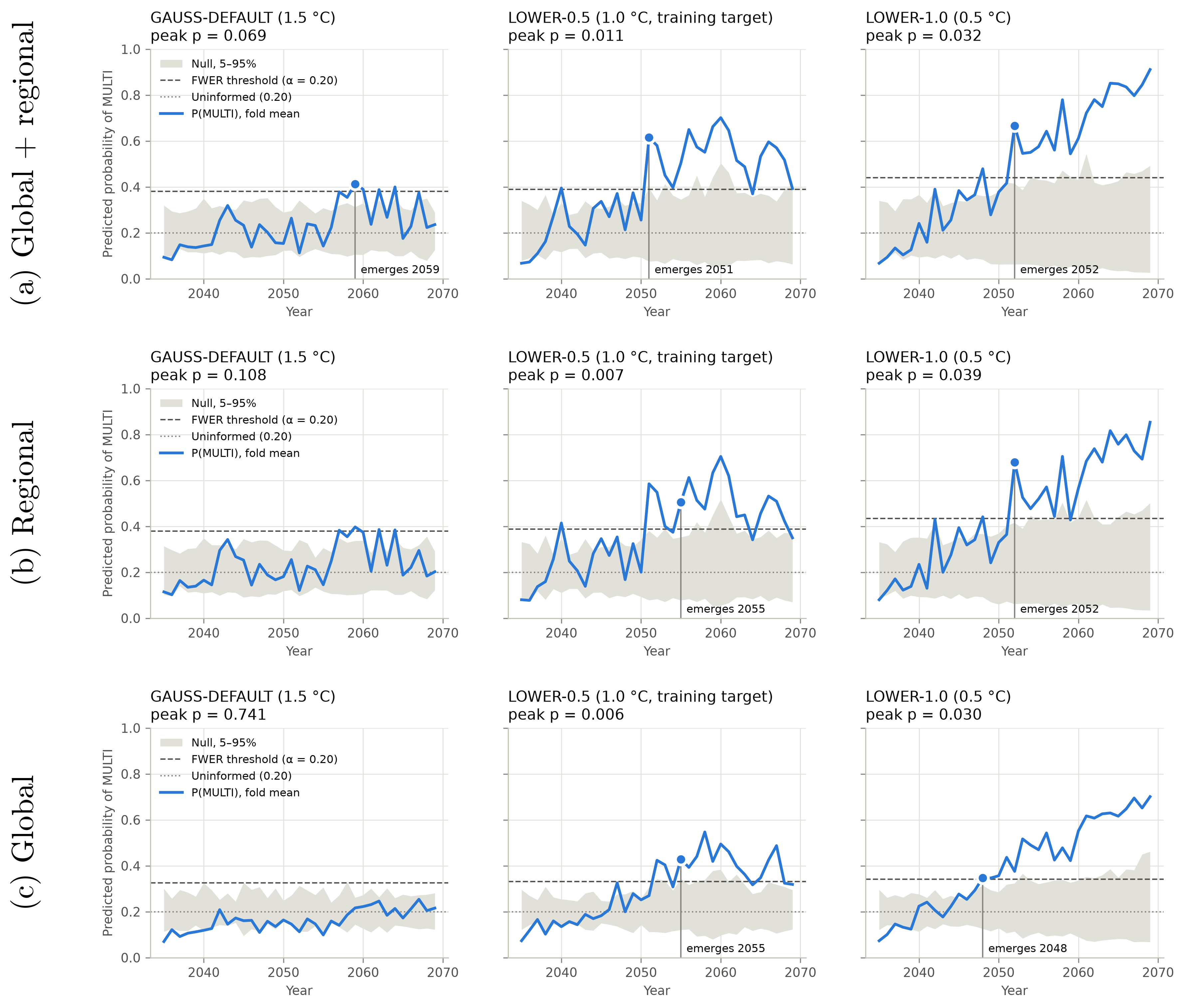}
\caption{Probability assigned to the multi-latitude strategy (MULTI) for held-out simulations at 1.5$^\circ$C (GAUSS-DEFAULT), the 1.0$^\circ$C training target (LOWER-0.5), and 0.5$^\circ$C (LOWER-1.0), from left to right, using global and regional (a), regional (b), and global (c) features. Shading shows the 5--95\% permutation null, the dashed line depicts the FWER threshold, and the dotted line indicates chance (0.20). Panel titles include uncorrected peak $p$-values. Emergence years are shown regardless of the significance of the peak $p$-value.
}
\label{fig:generalization_prob}
\end{figure}

\begin{figure}[tbp]
\centering
\includegraphics[width=\textwidth]{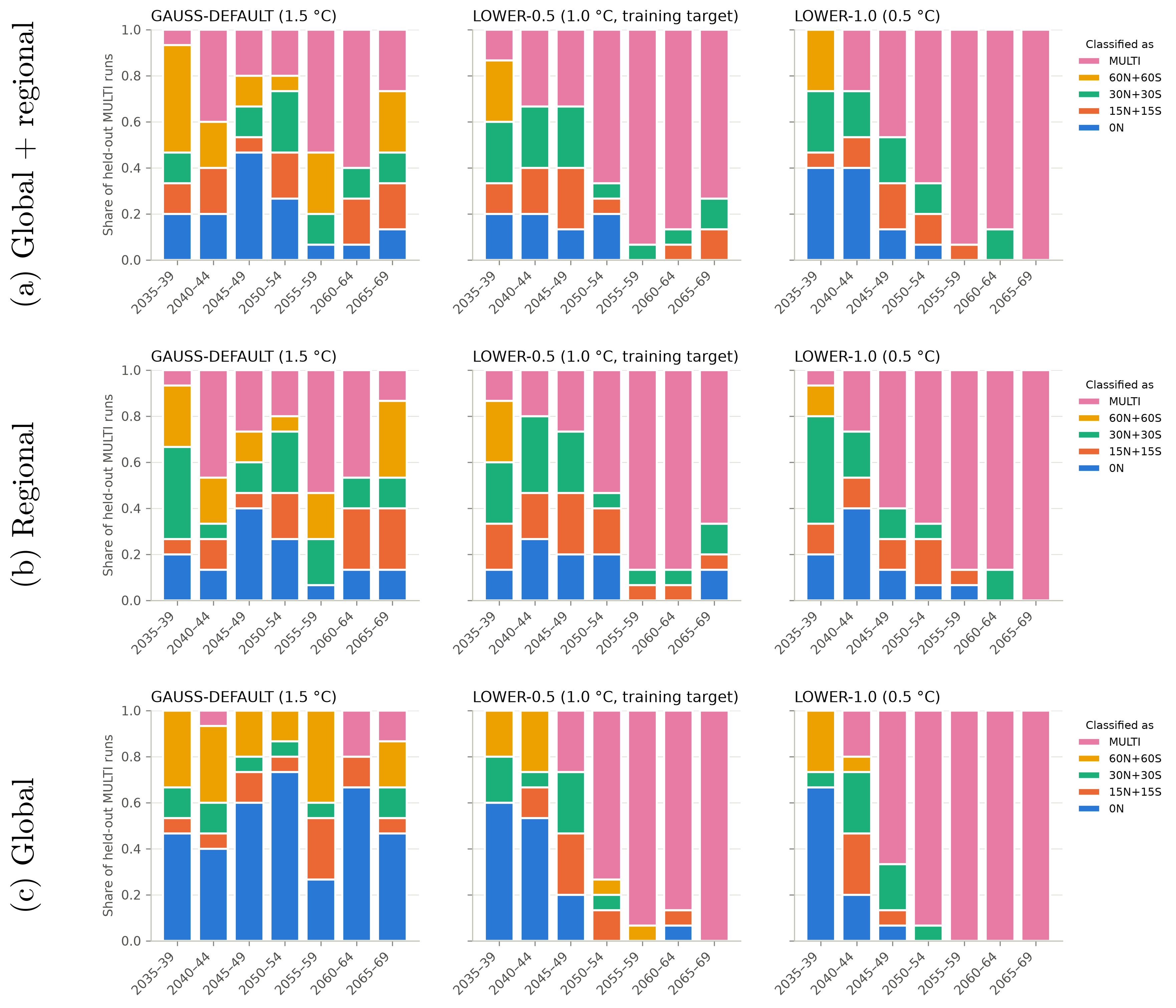}
\caption{Classification of the held-out multi-latitude (MULTI) simulations
at 1.5$^\circ$C (GAUSS-DEFAULT), the 1.0$^\circ$C training target
(LOWER-0.5), and 0.5$^\circ$C (LOWER-1.0), from left to right, using
global and regional (a), regional (b), and global (c) features. Bars show what the MULTI simulations are classified as in each five-year period.}
\label{fig:generalization_class}
\end{figure}

Figures~\ref{fig:generalization_prob} and~\ref{fig:generalization_class} show how well the classifiers perform at recognizing the multi-latitude strategy in simulations they were not trained on. The classifiers appear to get better at recognizing the multi-latitude strategy as the temperature target is lowered. A lower temperature target necessitates more cooling, and therefore more aerosol injection and a stronger signal. As shown by Visioni et al. \cite{visioni2023}, various surface impacts due to SAI scale approximately linearly with the amount of cooling, which suggests that the discriminative signal of the strategy is stronger at lower targets. For all three feature sets, the peak probability assigned to the multi-latitude strategy is lowest at the 1.5$^\circ$C target (0.25--0.41), intermediate at the 1.0$^\circ$C training target (0.55--0.70), and highest at the 0.5$^\circ$C target (0.70--0.91) (Figure~\ref{fig:generalization_prob}). The 0.5$^\circ$C results are significant for all three feature sets (uncorrected $p \le 0.039$), while none of the 1.5$^\circ$C results ($p \ge 0.069$) are significant. The strategies predicted by the classifiers show the same pattern (Figure~\ref{fig:generalization_class}). Starting in the 2050s, most 0.5$^\circ$C simulations are correctly classified as multi-latitude, while the 1.5$^\circ$C simulations are split among several strategies and less often classified as multi-latitude.

\section{Discussion}

Our work is closest to that of Labe et al. \cite{labe2023}, who used the same model, background scenario, and deployment year (CESM2(WACCM6), SSP2-4.5, 2035), to train a logistic-regression classifier with ensemble-member hold-out to predict on a yearly basis. Labe et al. sought to answer a binary question, predicting whether SAI had been deployed or not, using the ARISE 1.5$^\circ$C multi-latitude strategy \cite{richter2022}. They found the discriminative signal large enough that the model trained on global-land temperature alone was able to distinguish between the SAI and no-SAI scenarios perfectly after about the first five years (92.6\% over the full 2035–2069 period). Identifying which of the five strategies is deployed is a smaller-signal problem, since the GAUSS feedback controller keeps each strategy at the same global-mean temperature target. In our study, distinguishability emerges only in the mid-to-late 2040s -- about 8–14 years after deployment.

Our slower emergence period reflects the fact that all strategies have roughly the same climate goals, is a function of the more stringent family-wise permutation test, and may also reflect the size of the ensembles. We extend the previously discussed binary detection frameworks from the literature to the multi-class problem of determining which of several feedback-controlled strategies is deployed. We also introduce an emergence criterion that combines a family-wise permutation threshold with a calibrated two-year persistence requirement to identify the first year that distinguishability could be said to have emerged.

The above results are qualified by several limitations. The most important limitation is the size of the ensemble. With three members per strategy, the permutation reference set contains $(5!)^{3}\approx1.7\times10^{6}$ relabellings, but only $\approx1.4\times10^{4}$ are statistically distinct once permutations that are a uniform relabelling of the class labels are removed, resulting in a smallest attainable $p$-value of $\approx10^{-4}$. This floor restricts our statistical analysis in two ways. First, it constrains the significance of any individual model: the two most separable feature sets are already within an order of magnitude of the $p$-value floor (Table~\ref{tab:headline}), making it harder for our methodology to determine whether one model captures more discriminative power than the other. Second, the approach has little room for multiple-comparison FWER correction, since sharing a fixed significance budget across many tests drives the required threshold toward the floor, though this likely does not materially affect the relatively small model families considered in this study. 

Separately, the small size of the ensemble limits statistical power. A real but modest discriminative signal may not clear the significance threshold for two consecutive years. Therefore, the emergence years may be delayed by the limited ensemble size, and some single-variable models that do not survive correction may still contain useful discriminative signal. However, since most models provide no more than three ensemble members for their Coupled Model Intercomparison Project (CMIP)-style simulations, we believe our results are relevant for many other CMIP-related applications.

All our models are trained and tested on data generated using a single Earth system model (CESM2(WACCM6)). Since Earth system models differ substantially in their handling of atmosphere, ocean, and ice systems \cite{henry2023}, the emergence years and our discussion of the most discriminative variables and regions may be model-dependent. Internal variability of models has been shown to result in differences between modeled and observed climate even over several decades \cite{jain2023}. Since emergence year depends on the size of the distinguishable signal relative to internal variability, emergence may differ when applied to observed real-world climate. Integrating observed climate variability from reanalyses into the framework presents a promising direction for future work. Another promising direction is applying the framework presented in the current study to the Geoengineering Model Intercomparison Project (GeoMIP) experiments proposed for CMIP7 \cite{visioni2026}, which will include multiple SAI implementations and marine cloud brightening (MCB) across models under a common 1.5$^\circ$C temperature target, to test whether our results hold across models and whether different types of intervention (e.g.\ SAI vs.\ MCB) can be distinguished accurately.

All our models were trained on a linear classifier (multinomial logistic regression) using annual area-averaged means, which may ignore nonlinear features that a convolutional network trained on maps at full spatial resolution may learn to exploit. The regions were also chosen a priori based on a non-exhaustive literature review rather than chosen a posteriori to maximize separability.

The main analysis tests one cooling target ($0.5\,^{\circ}$C below $1.5\,^{\circ}$C) and five hemispherically symmetric strategies, in the perspective of analysing a set of strategies with similar aims and with a ``balanced'' intervention. Lower temperature targets and hemispherically asymmetric strategies may allow earlier emergence. The generalization tests were performed only on the multi-latitude strategy since the other strategies were not simulated at the other targets. Our generalization tests, therefore, measure whether the multi-latitude strategy is still detectable, not whether the five strategies remain distinguishable at different targets. The $p$-values in the generalization tests were also not corrected for multiple comparisons.

Finally, the run-based emergence-year metric is dependent on the selection of the significance band ($\alpha=0.20$) and the run-length calibration
($\alpha_{\mathrm{run}}=0.05$). These choices were not varied exhaustively. The regularisation strength was varied separately (Table~\ref{tab:csensitivity}), and the emergence year was shown to be sensitive to this choice.

\section{Conclusion}

This study asks whether the five feedback-controlled GAUSS injection strategies are distinguishable when trained and tested on annual regional and global means of surface temperature, precipitation, and polar sea-ice fraction. We also attempt to identify the year in which the discriminative signal becomes statistically detectable. We use a multinomial logistic-regression classifier that is tested using leave-one-member-out cross-validation against a restricted-permutation null. The strategies become separable in the mid-to-late 2040s, about 8--14 years after deployment and towards the end of the ten-year ramp-down to the 1.0$^\circ$C target, with peak cross-validated accuracy of $\sim 0.87$. In our study, the best performing model used both global and regional feature sets for all three variables. Regional temperature and precipitation models emerge later than models using both global and regional features, and fail the family-wise error correction. Since the feedback controller targets a common global-mean temperature, the global temperature feature is generally uninformative and the model trained on it is the worst performing headline model. Our results improve as more features are added at the global and regional resolutions, suggesting the discriminative signal is small and distributed across features.

The emergence reported in this study is later than that reported in previous literature aiming to detect whether SAI is deployed \cite{labe2023}, or to distinguish between uncontrolled unilateral injections \cite{dong2026}. This is expected given that the strategies all target the same global-mean temperature using a shared feedback controller, combined with a rigorous significance test. While the reported emergence years should be read as conditional on this model, ensemble size, and deployment schedule, our results suggest that given surface climate observed only at regional, annual resolution, roughly a decade or more of data is needed before a single year of data can identify with confidence which coordinated strategy has been deployed, but that the integration of a diverse set of monitored climate features can help shorten detection times. We also propose a concrete methodology to quantify emergence, combining a family-wise threshold derived using a restricted-permutation null distribution with a calibrated persistence requirement (two years in our main analysis). In the generalization tests, the classifiers perform better at recognizing the multi-latitude strategy as the temperature target is lowered. Classification of strategies may therefore depend on the temperature target used for training.

\data{{\raggedright
Output from the GAUSS simulations \cite{zhang2024} is
available from the corresponding author upon request; further information is available at
\url{https://climate-engineering.mae.cornell.edu/gauss-geoengineering-assessment-across-uncertainty-scenarios-and-strategies}. The code used to perform the analysis and produce the figures is available at \url{https://github.com/astrocharan/sai-classification/}.
\par}}

\end{document}